\documentclass[twocolumn]{aastex63}

\usepackage{graphicx}
\usepackage{amsmath}	
\usepackage{amssymb}
\usepackage{booktabs}
\usepackage{enumerate}
\usepackage{dirtytalk}
\usepackage{float}
\usepackage{enumitem}
\usepackage{booktabs}
\usepackage{lipsum} 
\usepackage{float}
\usepackage{xcolor}
\usepackage{multirow}
\usepackage{makecell}
\usepackage{blindtext}

\graphicspath{{./}{figures/}}

\submitjournal{AJ}
\accepted{December, 2022}

\shorttitle{On the Application of Bayesian Leave-One-Out Cross-Validation}
\shortauthors{Welbanks, McGill, Line \& Madhusudhan}

\begin{document}

\title{On the Application of Bayesian Leave-One-Out Cross-Validation to Exoplanet Atmospheric Analysis}

\author[0000-0003-0156-4564]{Luis Welbanks}
\thanks{NHFP Sagan Fellow}
\email{luis.welbanks@asu.edu}
\affil{School of Earth \& Space Exploration, Arizona State University, Tempe, AZ, 85257, USA}

\author[0000-0002-1052-6749]{Peter McGill}
\affil{Department of Astronomy \& Astrophysics, University of California, Santa Cruz, CA, 95064,USA}

\author[0000-0002-2338-476X]{Michael Line}
\affil{School of Earth \& Space Exploration, Arizona State University, Tempe, AZ, 85257, USA}

\author[0000-0002-4869-000X]{Nikku Madhusudhan}
\affil{Institute of Astronomy, University of Cambridge, Madingley Road, Cambridge, CB3 0HA, UK}

\begin{abstract}
Over the last decade exoplanetary transmission spectra have yielded an unprecedented understanding about the physical and chemical nature of planets outside our solar system. Physical and chemical knowledge is mainly extracted via fitting competing models to spectroscopic data, based on some goodness-of-fit metric. However, current employed metrics shed little light on how exactly a given model is failing at the individual data point level and where it could be improved. As the quality of our data and complexity of our models increases, there is an urgent need to better understand which observations are driving our model interpretations. Here we present the application of Bayesian leave-one-out cross validation to assess the performance of exoplanet atmospheric models and compute the expected log pointwise predictive density (elpd$_\text{LOO}$). elpd$_\text{LOO}$ estimates the out of sample predictive accuracy of an atmospheric model at data point resolution providing interpretable model criticism.  We introduce and demonstrate this method on synthetic HST transmission spectra of a hot Jupiter. We apply elpd$_\text{LOO}$ to interpret current observations of HAT-P-41b and assess the reliability of recent inferences of H$^-$ in its atmosphere. We find that previous detections of H$^{-}$ are dependent solely on a single data point. This new metric for exoplanetary retrievals complements and expands our repertoire of tools to better understand the limits of our models and data. elpd$_\text{LOO}$ provides the means to interrogate models at the single data point level, a prerequisite for robustly interpreting the imminent wealth of spectroscopic information coming from JWST. 

\end{abstract}


\section{Introduction}
\label{sec:intro}

Our understanding of exoplanet atmospheres has fundamentally changed in the last decade due to the plethora of space- and ground-based spectroscopic observations of transiting exoplanets from facilities such as the Hubble Space Telescope \citep[HST, e.g.,][]{Deming2013, McCullough2014, Sing2016}, the Spitzer Space Telescope \citep[e.g.,][]{Desert2011a, Pont2013, Stevenson2017}, the Very Large Telescope \citep[VLT, e.g., ][]{Nikolov2016, Sedaghati2017}, among others \citep[e.g.,][]{Chen2017, Rackham2017, Chen2018, Espinoza2019, vonEssen2019}. In particular, the transmission spectra of exoplanets, which encode the apparent change in planetary size as a function of wavelength, have provided us with important constraints on the chemical composition of exoplanet atmospheres \citep[e.g., ][]{Charbonneau2002, Deming2013, Sing2016}, information regarding the prevalence of clouds and hazes \citep[e.g.,][]{Pont2008, Nikolov2016, Kreidberg2014a, Benneke2019a}, and other atmospheric properties at the day-night terminator of the planet \citep[see e.g., ][for a review]{Madhusudhan2019}. 

The interpretation of these observations is routinely performed by means of model fitting and comparison techniques, i.e., atmospheric retrievals. Currently, atmospheric retrievals using Bayesian inference methods are widely employed to obtain parameter estimates from transit spectra \citep[see e.g.,][for a review]{Madhusudhan2018}. Through comparing competing atmospheric models, multiple chemical species have been identified and proposed as the cause of observed absorption features \citep[e.g.,][]{Nikolov2016, Sedaghati2017, Chen2017}. The retrieved constraints have enabled initial results on the abundances of these chemical species \citep[e.g.,][]{Kreidberg2014b, Barstow2017, MacDonald2017a}, as well as comparative studies searching for trends in the composition of exoplanet atmospheres \citep[e.g.,][]{Madhusudhan2014a,Pinhas2018, Fisher2018, Welbanks2019b}. Fundamentally, the conclusions drawn from exoplanetary atmospheric retrievals always depend on a comparison between competing models.

However, the inferences and conclusions made using these techniques can be conflicting. For instance, while a chemical species may be inferred using a specific data set and modeling strategy \citep[e.g.,][]{Sedaghati2017, Chen2017}, observations of the same planet obtained with different facilities and interpreted with different models, may not lead to the same conclusions \citep[e.g.,][]{Espinoza2019, Spake2021}. Likewise, the resulting best-fit solutions can sometimes be incompatible with the observations according to some goodness-of-fit metrics \citep[e.g.,][]{MacDonald2019, Colon2020}. Interpretations of the same observations using different models and model fitting techniques have also led to opposing conclusions \citep[e.g.,][]{Sing2016, Barstow2017, Pinhas2019}. Overall, these apparent disagreements highlight the need for a thorough understanding of the limits of our models and the sensitivity of our inferences on the spectroscopic observations used.

It is desirable to have a method to assess the robustness of model inferences and their sensitivity to individual observations. However, existing model assessment metrics such as the Bayesian evidence, $\chi^2$, p-value, or Bayesian Information Criterion (BIC), provide a single global summary value for the performance of the model conditioned on the entire dataset \citep[e.g.,][]{Madhusudhan2009, Benneke2013, Colon2020, Alderson2022}, making it difficult to shed light on which specific observations affect the model performance. Furthermore, these commonly used metrics can be hard to interpret as their assessment of the model performance may be at odds with each other \citep[e.g.,][]{Colon2020, Lewis2020, Sheppard2021}. Besides these metrics, some studies have used an information content based approach \citep[i.e., quantifying the change in the state of knowledge by making a measurement, e.g.,][]{Line2012, Batalha2017b} to estimate the wavelengths and the precisions required to better inform the parameters in atmospheric models of transmission spectra. However, these efforts have been limited to simple atmospheric models with a small number of parameters. Additionally, the information content based approach has yet to be successfully implemented as a tool for model comparison and model assessment in atmospheric retrievals of transmission spectra. 

There is a need for complementary model performance metrics that can assess performance both at the dataset and individual data point level. These metrics must take into account the sources of uncertainty in our models, and be generally applicable to non-linear atmospheric models of increasing complexity. Crucially, for their successful application to atmospheric retrievals, these metrics must be computationally inexpensive. The advent of such tools is paramount upon the imminent data release from the James Webb Space Telescope (JWST) at wavelengths inaccessible by existing facilities and at unprecedented precisions \citep[e.g.,][]{Greene2016}. Such a model performance metric is also key to ensure reliable and robust inferences and conclusions from current and upcoming observations. 

In this paper, we explore the application of the expected log pointwise predictive density estimated by Bayesian leave-one-out cross-validation (elpd$_\text{LOO}$) as an interpretable model selection, comparison, and criticism tool. While elpd$_\text{LOO}$ is starting to be applied in some areas of astronomy \citep[e.g.,][]{Morris2021,Meier2022, McGill2022, Neil2022}, has yet to be used in the context of atmospheric retrievals of transmission spectra. elpd$_\text{LOO}$ \citep[e.g.,][]{Vehtari2012} is a metric that quantifies the out of sample predictive accuracy of a model, e.g., how well does the model predict unseen data. This Bayesian metric can provide an indication of model performance at the data point or dataset level. Besides single-model assessment, elpd$_\text{LOO}$ is a powerful tool for model comparison as it can highlight the relative performance of two models at the data point resolution. Using the Pareto Smoothed Importance Sampling \citep[PSIS;][]{Vehtari2015} approximation for elpd$_\text{LOO}$ we demonstrate that this metric is computationally feasible for exoplanet atmospheric models. 

We begin by summarizing the method of atmospheric retrievals as well as the advantages and disadvantages of the statistical metrics currently employed for model selection in Section \ref{sec:retrieval_and_metrics}. In Section \ref{sec:loo} we introduce the concept of Bayesian cross validation and the algorithm to compute elpd$_\text{LOO}$. We explain our atmospheric retrieval setup in Section \ref{sec:retrieval_setup} and in Section \ref{sec:simulated_data} we demonstrate the power of elpd$_\text{LOO}$ in atmospheric retrievals of transmission spectra by testing elpd$_\text{LOO}$ on synthetic observations of a hot Jupiter. Then, we use elpd$_\text{LOO}$ to assess the robustness of recent inferences of the hydrogen anion H$^-$ in the atmosphere of the hot Jupiter HAT-P-41b in Section \ref{sec:hatp41b}. We conclude by summarizing our findings and exploring future research directions in Section \ref{sec:summary_discussion}.

\section{Model Assessment in Exoplanet Atmospheric Retrievals}
\label{sec:retrieval_and_metrics}

Atmospheric retrieval frameworks are powerful tools to infer the atmospheric properties of an exoplanet from its spectrum. At their core these tools couple an atmospheric model describing the physical state of the planetary atmosphere, a prior distribution over the possible values of the model parameters, and a parameter estimation scheme. Although a wide range of parameter estimation schemes have been used \citep[see e.g., ][for a review]{Madhusudhan2018}, nested sampling \citep[e.g.,][]{Skilling2006, Feroz2009, Buchner2014} prevails in the literature.

\subsection{Frequentist Hypothesis Testing}

Fit quality metrics like $\chi^{2}$ and corresponding p-values are commonly used in retrieval studies to assess whether or not a model adequately explains the data \citep[e.g.,][]{Colon2020, Lewis2020, Sheppard2021, Alderson2022}. Studies generally compute the reduced $\chi^{2}$, $\chi^{2}_\text{red}$ = $\chi^{2}_{\mathcal{M}}/\nu$ where $\nu$ is the degrees of freedom, and values greater than one are considered ``poor'' fits, while values smaller than one are considered an overfit \citep[][]{Andrae2010}. Similarly, the p-value is interpreted as the probability that random noise can result in the observed data given that the model being tested is true, with some studies considering p-values $<10^{-3}$ indicative of ``poor'' fits \citep[e.g.,][]{Colon2020}. 

However, these fit quality metrics have two main pitfalls within the context of atmospheric retrievals. First, the atmospheric models used in exoplanetary retrievals are not linear in their parameters. As a result, estimating the true number of degrees of freedom of a given model to compute $\chi^{2}_\text{red}$ is not possible. Second, current studies ignore the uncertainty in the estimate of $\chi^{2}$ which in practice can be difficult to estimate. These limitations affect estimates of the p-value as this metric relies on the assumption that $\chi^{2}$ statistics are appropriate for the model being tested. Furthermore, the p-value penalizes complex models with a large number of parameters, likely resulting in a preference for simpler and possibly incomplete models. 

\newpage
\subsection{Bayesian Hypothesis Testing}

More recently, model comparison/selection using the Bayesian Evidence\footnote{The Bayesian Information Criterion (BIC), which is an approximation to the Bayesian Evidence, has also been used in the field to perform model comparison/selection \citep[e.g.,][]{Sotzen2020, Spake2021, Alderson2022}}\citep[e.g.,][]{Benneke2013,Sedaghati2017,Welbanks2021} has become customary due to nested sampling techniques which permit its computation \citep[e.g.,][]{Skilling2006,Feroz2009,Buchner2014}. This metric is computed by taking the ratio of evidences (also know as the Bayes' factor) between two models. Additionally, the difference in evidence between models has been mapped to a commonly used `sigma' scale and introduced to the field of exoplanetary retrievals by \citet{Benneke2013}. Besides its straightforward computation, Bayesian model evidence has the appealing property of penalizing model complexity that is not supported by the data effectively applying Occam's razor to model selection.

Bayesian metrics allow the incorporation of prior information about a parameter when calculating the probability distribution of interest. Under pathological circumstances, this key feature could be misused by a practitioner. For instance, a different choice of prior, e.g., to one more or less informative, will result in a different Bayesian evidence due to the associated change in prior space, and may lead to a different interpretation regarding the model preference. This presents a challenge in the context of exoplanet atmosphere because commonly, the priors for the model parameters (e.g., chemical abundances, P-T profile parameters, etc.) have somewhat arbitrary boundaries and vary significantly different between studies. For instance, changing the priors on a given chemical species (e.g., H$_2$O) from uniform in log-space \citep[e.g., -1 to -12, see e.g., ][]{Welbanks2019a} to a prior motivated by the species' thermo-chemical equilibrium expectations (e.g., logarithmic from -2 to -4) would significantly reduce the prior space and could result in a very different model evidence and associated interpretation, without necessarily changing any posterior inferences. Therefore, improper use of these Bayesian metrics through e.g., inappropriate prior choices, can propagate through to the Bayes factor, and potentially affect the reported detection significance.
 
\section{Leave-One-Out Cross-Validation}
\label{sec:loo}

The model assessment metrics described in Section \ref{sec:retrieval_and_metrics} provide a single value shedding little light on exactly where and how each model is failing or performing well. From this single number, it is non-trivial to estimate whether the models would perform better or worse at other wavelengths, or to estimate which parts of the data are driving our inferences. To overcome these challenges, we propose the application of Bayesian Leave-one-out cross-validation to compute the expected log pointwise predictive density (elpd$_\text{LOO}$) to aid interpretation of atmospheric retrievals. In the rest of this section we closely follow \citet{Vehtari2017}, and outline the theory behind elpd$_\text{LOO}$. 

Consider an atmospheric model $\mathcal{M}$, which is being fit to a spectroscopic data set $\mathcal{D}$. Here $\mathcal{D}=\{\mathcal{D}_{i}\}_{i=1}^{N}$ is the set of $N$ spectroscopic data points where, $\mathcal{D}_{i}=\{\Delta_{i}, \sigma_{\Delta_i}\}$ comprised of a transit depth ($\Delta_{i}$), and transit depth error ($\sigma_{\Delta_i}$) pair, at a wavelength $\lambda_{i}$. Then, using Bayes' theorem the posterior distribution for each of the model parameters ($\vec{\theta}$) for a given model ($\mathcal{M}$) is

\begin{equation}
    p(\vec{\theta} | \mathcal{D}, \mathcal{M}) = \frac{p(\mathcal{D}|\vec{\theta},\mathcal{M})p(\vec{\theta}|\mathcal{M})}{p(\mathcal{D}|\mathcal{M})}.
    \label{eq:bayes_rule}
\end{equation}
\noindent Here, $p(\mathcal{D}|\vec{\theta},\mathcal{M})$ is the likelihood and represents the probability of observing the data ($\mathcal{D}$) given a specific set of model parameters ($\vec{\theta}$). The likelihood is multiplied by the prior distribution $p(\vec{\theta}|\mathcal{M})$ and divided by the Bayesian model evidence $p(\mathcal{D}|\mathcal{M})$. A retrieval uses the likelihood and prior to compute estimates for the Bayesian model evidence and posterior distributions for the model's parameters. 

To complement existing model selection and criticism, we introduce the leave-one-out expected log posterior predictive density elpd$_\text{LOO}$ obtained by Bayesian cross-validation. Bayesian leave-one-out cross-validation is one way to estimate the out of sample predictive accuracy of a model \citep[i.e., the expected log pointwise predictive density, see e.g.,][for a review]{gelman2014}. Bayesian leave-one-out cross-validation works by fitting a given model $\mathcal{M}$ to $\mathcal{D}_{-i}$, the full data set ($\mathcal{D}$) with the $i$th data point, $\mathcal{D}_{i}$, left out.  The posterior distribution of the fit it used to assess how well the left out data point is predicted. This procedure is then repeated for each data point in turn. Ultimately, this allows each data point to be scored on how well it is predicted by a given model conditioned on the rest of the dataset. These data point scores can then be compared pairwise between competing models or totaled over the data set to give an indication of overall model performance. The elpd$_\text{LOO}$ for $\mathcal{D}_{i}$ is the expected log posterior predictive density,

\begin{equation}
\text{elpd}_{\text{LOO}, i, \mathcal{M}} = \log p(\mathcal{D}_{i}|\mathcal{D}_{-i}, \mathcal{M}).
\label{eq:loo_score}
\end{equation}

The posterior predictive density can be calculated by taking the expectation of the likelihood of the $\mathcal{D}_{i}$ with respect to the posterior of the model fitted to $\mathcal{D}_{-i}$,

\begin{equation}
    p(\mathcal{D}_{i}|\mathcal{D}_{-i}, \mathcal{M}) = \int p(\mathcal{D}_{i}|\vec{\theta},\mathcal{M})p(\vec{\theta}|\mathcal{D}_{-i},\mathcal{M})d\vec{\theta}.
    \label{eq:elpd}
\end{equation}

The elpd$_\text{LOO}$ score quantifies how well the model fit to the data (i.e., trained model under the chosen prior), performs on unseen data where each data point is left out in turn. On the other hand, the Bayesian evidence quantifies the probability of the data agreeing with the specified model under the chosen prior. Therefore, they are fundamentally different quantities that when used together can enrich our understanding of our models and their ability to explain the data.

In practice, if we have $S$ samples from the posterior distribution obtained by fitting the model to $\mathcal{D}_{-i}$ ($\{\hat{\theta}^{s}_{-i}\}_{s=1}^{S}$), we can estimate the posterior predicted density of $\mathcal{D}_{i}$ with,

\begin{equation}
    p(\mathcal{D}_{i}|\mathcal{D}_{-i},\mathcal{M})= \frac{1}{S}\sum_{s=1}^{S}p(\mathcal{D}_{i}|\hat{\theta}^{s}_{-i}, \mathcal{M}).
    \label{eq:elpd_exact}
\end{equation}

\noindent All data point scores can be totaled over the full data set to give a indication of overall model performance,

\begin{equation}
    \text{elpd}_{\text{LOO},\mathcal{M}} = \sum_{i=1}^{N}\text{elpd}_{\text{LOO}, i, \mathcal{M}}.
    \label{eq:total_loo}
\end{equation}

\noindent Two competing model's $(\mathcal{M}_{1},\mathcal{M}_{2})$ overall performance can be computed by taking the difference in elpd score with,

\begin{equation}
    \Delta\text{elpd}_{\text{LOO},\mathcal{M}_{1},\mathcal{M}_{2}} = \text{elpd}_{\text{LOO},\mathcal{M}_{1}} - \text{elpd}_{\text{LOO},\mathcal{M}_{2}}.
    \label{eq:loo_diff}
\end{equation}

\noindent Here, a positive $\Delta\text{elpd}_{\text{LOO},\mathcal{M}_{1},\mathcal{M}_{2}}$ would mean that $\mathcal{M}_{1}$ has the better out of sample predictive performance. 

In principle, calculating the exact elpd$_\text{LOO}$ score would require fitting the model to infinite data. Therefore, the assumption here is that the $N$ data points used to calculate the elpd$_\text{LOO}$ score come from a larger population making it possible to calculate their standard error (SE). The SE of the estimated elpd$_\text{LOO}$ score is

\begin{equation}
    \text{SE} (\text{elpd}_{\text{LOO},\mathcal{M}}) = \sqrt{N\text{V}_{i=1}^{N}(\text{elpd}_{\text{LOO},i, \mathcal{M}})},
    \label{eq:se_loo}
\end{equation}

\noindent while the SE for the difference between two scores is given by,

\begin{align}
\begin{split}
    &\text{SE} (\Delta\text{elpd}_{\text{LOO},\mathcal{M}_{1},\mathcal{M}_{2}})
    \\&=\sqrt{N\text{V}_{i=1}^{N}(\text{elpd}_{\text{LOO},i, \mathcal{M}_{1}}-\text{elpd}_{\text{LOO}, i, \mathcal{M}_{2}})}.
\end{split}
    \label{eq:se_diff}
\end{align}
Here, $V$ is the variance operator. 

Naive computation of all $\text{elpd}_{\text{LOO}, i, \mathcal{M}}$ terms for a given model would be computationally expensive. This is because each term requires a full Bayesian refit of the model to obtain posterior samples. Overall, $N$ full Bayesian fits would be required. In the case of exoplanet atmospheric modeling, $N\approx100-1000$ and one Bayesian refit, typically performed with nested sampling, can take $\gtrsim 27$ Central Processing Unit (CPU) hours \citep[e.g.,][]{Zhang2020}. This required computation is clearly prohibitive, therefore, we have to turn to the Pareto Smoothed Importance Sampling \citep[PSIS;][]{Vehtari2015} approximation outlined in  \citep{Vehtari2017} to calculate $\text{elpd}_{\text{LOO}, i, \mathcal{M}}$\footnote{We used the code available at \url{https://github.com/avehtari/PSIS} for the PSIS approximation. The relevant PSIS methods are also available in the Arviz python package \citep{arviz_2019}.}. 

We defer a detailed explanation of the PSIS approximation to Appendix \ref{app:psis_method}, but briefly outline the method here following \cite{Vehtari2015} and \cite{Vehtari2017}. PSIS allows us to approximately compute all $\text{elpd}_{\text{LOO}, i, \mathcal{M}}$ for a model without having to refit the model $N$ times. When using the PSIS approximation, the model is fit to the whole data once (e.g., one retrieval) and the posterior samples from this single inference are re-weighted using importance sampling to approximate the effect of leaving each data point out in turn. Instead of raw importance weights being using in the approximation, the distribution of importance weights is fitted with a Pareto distribution \citep{zhang2009} and a set of smoothed importance weights is drawn from this distribution and used. This is called PSIS and it serves two purposes. Firstly, PSIS acts to stabilize the importance sampling approximation by removing extreme weights that can dominate and make the importance sampling approximation unreliable \citep{Vehtari2015}. Secondly, the fitted shape parameter of the Pareto distribution, $k$, traces the number of finite moments of the importance weights distribution and acts as diagnostics on the reliability of the importance sampling approximation. For LOO, $k$ identifies cases when the posterior distribution changes significantly upon removing a data point and, therefore, the importance sampling approximation would be unreliable and should not be used. \citet{Vehtari2015} determined empirically that PSIS estimates were reliable for $k\lessapprox0.7$. In the case of LOO, the PSIS approximation is used for each data point where $k<0.7$. For data points where $k\geq0.7$, the PSIS approximation is likely to be unreliable, so a full refit of the model is performed and used to calculate $\text{elpd}_{\text{LOO}, i, \mathcal{M}}$. Overall, this allows all $\text{elpd}_{\text{LOO}, i, \mathcal{M}}$ terms to be computed with less than $N$ refits of the model. For brevity, in what follows $\text{elpd}_{i, \mathcal{M}}\equiv\text{elpd}_{\text{LOO}, i, \mathcal{M}}$.

\subsection{Limitations and other considerations}

Bayesian leave-one-out cross-validation presents an additional tool for model criticism and comparison within the context of atmospheric retrievals. This tool allows us to compare two models on a per-data-point scale, but as with any other model comparison metric, it cannot be used in isolation to argue definitively for or against a detection of a spectral feature in an exoplanet atmosphere. The reliability of an inference can only be robustly established by considering multiple model comparison and criticism tools and metrics.

For instance, $\text{elpd}_{\text{LOO},\mathcal{M}}$ does not penalize for model complexity. Therefore, when trying to assess two models with varying degrees of complexity, Bayesian leave-one-out cross-validation could be used alongside metrics that have a built in Occam's razor such as the Bayesian evidence. Additionally, as is the case for existing analysis using the Bayesian evidence, when the set of possible models is large, model comparison with $\text{elpd}_{\text{LOO},\mathcal{M}}$ may result in a combinatorial explosion of possible pairwise combinations.

Additionally, the difference in $\text{elpd}_{\text{LOO},\mathcal{M}}$ score (Equation \ref{eq:loo_diff}) between different pairs of models can be difficult to interpret and has not been commonly mapped to a `sigma' scale as other model comparision metrics have. Moreover, the standard error for the difference between two scores (Equation \ref{eq:se_diff}) comes from the consideration that we are estimating the LOO predictive accuracy of a model using a finite sample of $N$ data points \citep[see e.g., ][]{Vehtari2017}. As such, the standard error estimate may be inaccurate if the number of observations is limited \citep[see e.g.,][for a discussion on the uncertainty in elpd estimates]{Sivula2020}. Therefore, we caution against using the difference of $\text{elpd}_{\text{LOO},\mathcal{M}}$ scores in units of standard errors as a proxy for sigma significance, but it can be useful to visualize the relative performance of different models. We further explore this and other future considerations in Section \ref{subsec:future}.

\section{Atmospheric Retrieval Setup}
\label{sec:retrieval_setup}

First, we demonstrate the use of elpd on synthetic HST STIS and HST WFC3 observations, the common observational setup in the pre-JWST era, to obtain intuition for the information that this metric can provide as well as demonstrate its conformance with prior expectations. Then, we apply this tool to the interpretation of current observations of the transiting hot Jupiter HAT-P-41b \citep{Hartman2012} obtained with HST WFC3 UVIS G280, HST WFC3 G141, and Spitzer IRAC from \citet{Wakeford2020} and \citet{Lewis2020} spanning the wavelength range 0.2 to 5.0~$\mu$m. Particularly, we assess the previous detection of H$^-$ absorption in the planet's atmosphere -- a result in significant discrepancy with thermo-chemical expectations. 

The observations are interpreted using Aurora \citep{Welbanks2021} with the atmospheric forward model following the standard prescription in \citet{Welbanks2019a, Welbanks2019b} and \citet{Welbanks2021}. The model atmosphere is divided into 100 layers equally spaced logarithmically from 10$^{-6}$-$10^{2}$~bar. The Bayesian inference is performed using the nested sampling algorithm \citep{Skilling2004} through MultiNest \citep{Feroz2009} and the Python package PyMultiNest \citep{Buchner2014}.

\subsection{Synthetic Observations}
\label{subsec:simulated_data}

We generate synthetic HST-STIS and HST-WFC3 synthetic observations assuming spectral resolutions and precisions comparable to existing observations \citep[e.g.,][]{Sing2016}, generally following the procedure described in \citet{Welbanks2022}. First, we generate spectra at a higher resolution of $R\sim13,000$ on average, from $0.3$--$2.0$~$\mu$m sampled line-by-line, and solving radiative transfer in transmission geometry. The resulting spectra is used to produce a binned spectrum following the methods outlined in \citet{Pinhas2018}. The resulting binned spectra has a resolution and precision of $R_{\rm HST-STIS}=20$ and $100$~ppm, respectively, for HST-STIS, and $R_{\rm HST-WFC3}=60$ and $50$~ppm for HST-WFC3. The synthetic observations include Gaussian scatter. 

The synthetic observations are generated for a planet with the same system parameters as HD~209458b \citep[e.g., $R_{\rm p} = 1.359 R_{\rm J}$, $M_{\rm p} = 0.685 M_{\rm J}$, and $R_{\mathrm{star}}=1.155 R_\odot$][]{Torres2008}. The model generating the synthetic data assumes a clear, H$_2$-rich, isothermal atmosphere at the equilibrium temperature of the planet $T_{\rm eq}=1450~K$, with solar abundances of H$_2$O, Na, and K \citep[i.e.,  $\log(X_{\text{H}_2\text{O}})=-3.3$, $\log(X_{\text{Na}})=-5.76$, $\log(X_{\text{K}})=-6.97$, e.g., ][]{Asplund2009}. We assume a reference pressure for the planetary radius of 10 bar. 

\subsection{Atmospheric Model}
\label{subsec:model_setup}

Our atmospheric model computes radiative transfer in a parallel-plane planet atmosphere in transmission geometry in hydrostatic equilibrium. The chemical absorbers considered in this work are H$_2$O \citep{Rothman2010}, Na \citep{Allard2019}, K \citep{Allard2016}, CrH \citep{Bauschlicher2001}, AlO \citep{Patrascu2015}, VO \citep{McKemmish2016}, bound-free H$^-$ \citep{John1988}, and H$_2$-H$_2$ and H$_2$-He collision induced absorption \citep[CIA;][]{Richard2012}. 

First, to test the implementation of elpd, we use a similar model to the one that generated the synthetic observations. Namely, the atmospheric model considers absorption due to H$_2$O, Na, and K, with the volume mixing ratio of each species as a free parameter. The Pressure-Temperature (P-T) profile is assumed to be isothermal, with 1 free parameter for the atmospheric temperature and 1 free parameter for the reference pressure ($P_{\rm ref. }$) at the assumed planet radius of $R_{\rm p} = 1.359 R_{\rm J}$. Instead of assuming a clear atmosphere, the model considers the presence of inhomogeneous clouds and hazes following the generalized parameterization in \citet{Welbanks2021} using one fraction for the combined effects of clouds and hazes. In total, this model has 9 free parameters: 3 chemical species, 1 isothermal temperature, 1 reference pressure, and 4 parameters for inhomogeneous clouds and hazes.

Second, we implement the same atmospheric model used by \citet{Lewis2020} to interpret the HAT-P-41b observations and used to infer a $\gtrsim2.6\sigma$ detection of H$^-$ (i.e., their ``minimal'' model setup). This atmospheric model has an isothermal, clear atmosphere with absorption due to H$_2$O, Na, CrH, AlO, VO, and H$^-$. Following \citet{Lewis2020}, we retrieve the planetary radius for a reference pressure at 10~bar. In total, this model has 8 parameters: 6 chemical species, 1 for the isothermal temperature of the atmosphere, and 1 for $R_{\rm p, ref.}$. We follow the retrieval setup from \citet{Lewis2020} to enable a direct comparison with their results suggesting the presence of significant H$^-$ opacity in the atmosphere of HAT-P-41b.

\section{Interrogating a known model}
\label{sec:simulated_data}

\begin{figure*}
\includegraphics[width=1.0\textwidth]{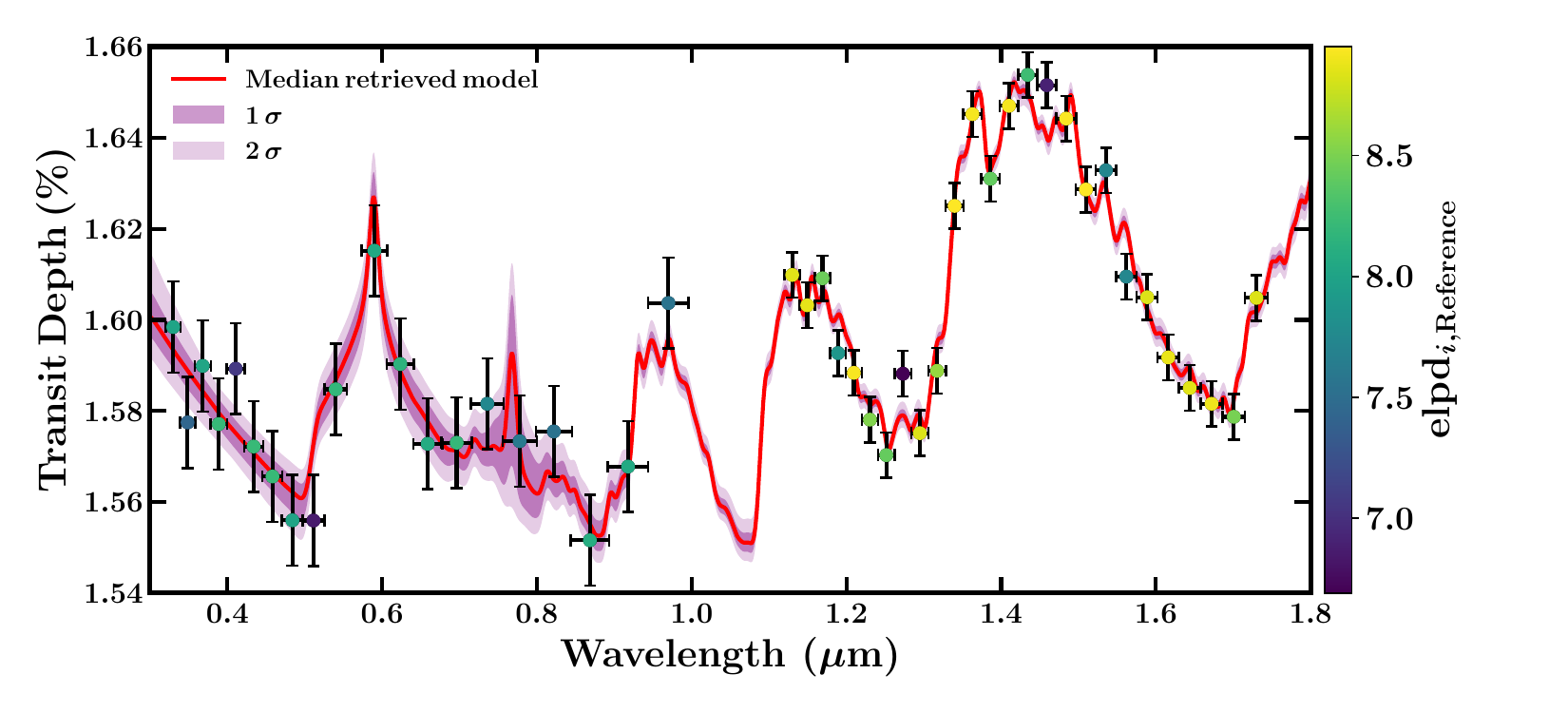}
\caption[]{Reference model fit and simulated data as described in Section \ref{sec:simulated_data}. Each of the data points are colored by their PSIS approximated elpd score ($\text{elpd}_{i,\text{Reference}}$). Data points with a lighter (more yellow) color have a higher elpd score and are better predicted by the model when left out.} \label{fig:synth_model_loo_score}
\end{figure*} 

Before applying elpd to real spectroscopic data, we first build intuition by computing this metric on the synthetic data described in Section \ref{subsec:simulated_data}. In this case we know the true underlying model that generated the data. By knowing the true model and the physical properties it represents, we can sense check the elpd score in the context of exoplanet atmospheric models, and build intuition on how to apply it to real data sets. In this section we use the atmospheric model explained in Section \ref{subsec:model_setup} considering 3 chemical absorbers, an isothermal atmosphere, an the possibility of inhomogeneous clouds and hazes.  In what follows we compute the elpd score using the PSIS approximation and found it to be accurate and significantly faster than the naive method of computing elpd saving $\approx4,000$ CPU hours of computation (see Appendix \ref{app:psis_validation}).

As commonly performed in the exoplanetary atmospheric retrieval literature, we compare a `reference model' with the highest degree of complexity, to a series of less complex models for which one or more parameters, or model components, have been removed. Here, our `reference model' is described in Section \ref{subsec:model_setup} and is similar to the model that generated the synthetic observations. Namely, our reference model considers absorption due to H$_2$O, Na, K, an isothermal P-T profile, and the possibility of inhomogeneous clouds and hazes. Then, we compare this reference model to four other models: 1) same as reference model but without H$_2$O absorption, 2) same as reference model but without Na absorption, 3) same as reference model but without K absorption, and 4) same as reference model but without the presence of inhomogenous clouds and hazes.

Performing the model comparison using the difference of the Bayesian evidence and converting it to a `detection significance' as commonly performed in the literature \citep[e.g.,][]{Benneke2013, MacDonald2017a, Welbanks2021} results in a model preference for H$_2$O of $27\sigma$, Na of $6\sigma$, K of $2\sigma$, while the inclusion of inhomogeneous clouds in the model is not preferred at a $2\sigma$ level. This comparison of the Bayesian evidence supports the known input of the model generating the data: the presence of H$_2$O, Na, and K. Additionally, the Bayesian evidence comparison indicates that the additional model complexity associated with including the presence of inhomogeneous clouds and hazes is not supported by the data. 

Despite the three chemical components being present in the true model generating the data, their associated `detection significances' are substantially different. This may be due to the fact that a large number of data points cover the broad H$_2$O absorption features while the main features due to Na and K are only encompassed by a few spectral bins. However, this information is not provided by the model comparison metric by itself. Additionally, the interpretation of a `detection significance' higher than $10\sigma$ \citep[i.e., a probability of 1-7$\times10^{-24}$, e.g.,][]{MacDonald2019} is difficult as it represents the relative preference between two models which may be equally incorrect.  Finally, the difference in the Bayesian evidence does not provide any information at the data-point (or spectral feature) resolution about the strengths or deficiencies of the model.

\begin{figure*}[t!]
\includegraphics[width=1.0\textwidth]{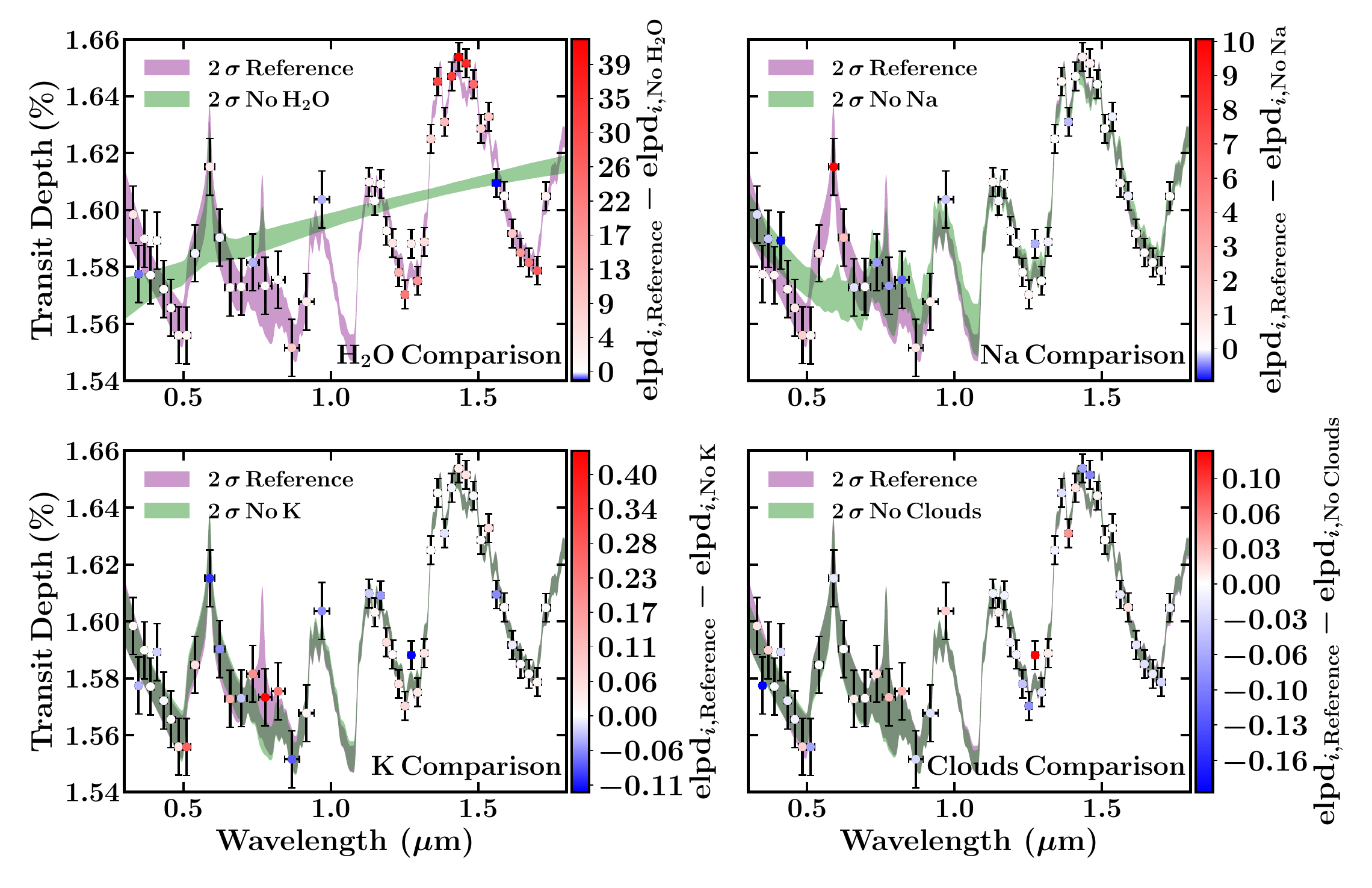}
\caption[]{Comparison between the reference model (purple) and simpler models (green) with shaded regions showing their retrieved 2$\sigma$ confidence intervals. The synthetic observations are color coded by their $\Delta$elpd score following Equation \ref{eq:loo_diff}, between the reference model and models without H$_2$O (top left), Na (top right), K (bottom left), and inhomogeneous clouds and hazes (bottom right). Data points with larger positive $\Delta$elpd scores (redder points) indicate that the reference model is better at explaining them. H$_2$O absorption preferentially explains data points at $\sim1.1$-$1.7\mu$m, Na absorption at $\sim0.6\mu$m, and K absorption at $\sim0.8\mu$m, in agreement with expectations. The increase in the predictive performance of the model (the numerical scale of the color map) is largest due to H$_2$O, followed by Na, and K; the inclusion of inhomogeneous clouds and hazes in the model does not improve the predictive performance of the model for this input cloud free atmosphere.} \label{fig:synth_model_loo_comparison}
\end{figure*} 

Detailed information about the model performance at the per-data-point level can be teased out using the elpd metric. Figure \ref{fig:synth_model_loo_score} shows the retrieved median model and the confidence intervals of the retrieval using the reference model on the synthetic observations. The synthetic observations have been color coded by their elpd score. The darker blue spectral points in Figure \ref{fig:synth_model_loo_score} have a lower elpd score than the bright yellow points, and are comparatively worse predicted by the atmospheric model, when left out of the fit. For instance, the elpd score suggests that the spectral point at $1.27\mu m$, which has the lowest elpd score of $6.7$, is the worst explained by the model as shown by the data fit, where it is far away from the model. 

The primary utility of elpd lies in model comparison. As such, Figure \ref{fig:synth_model_loo_comparison} shows the per-data-point difference in elpd score ($\Delta$elpd score) between the reference model and the less complex models used\footnote{The elpd scores for all models are calculated using the PSIS approximation. The Pareto $k$ values for all data points for all models are $<0.7$ indicating that the approximation is reliable.}. Additionally, Figure \ref{fig:synth_model_loo_comparison} shows the retrieved confidence interval for each of the models compared. The top left panel, showing the comparison of the models including and not including H$_2$O absorption, has the highest difference in elpd scores. The magnitude of the difference is then followed by the models comparing the presence of Na, then K, and finally the presence of inhomogeneous clouds and hazes.

\begin{figure}
\includegraphics[width=0.48\textwidth]{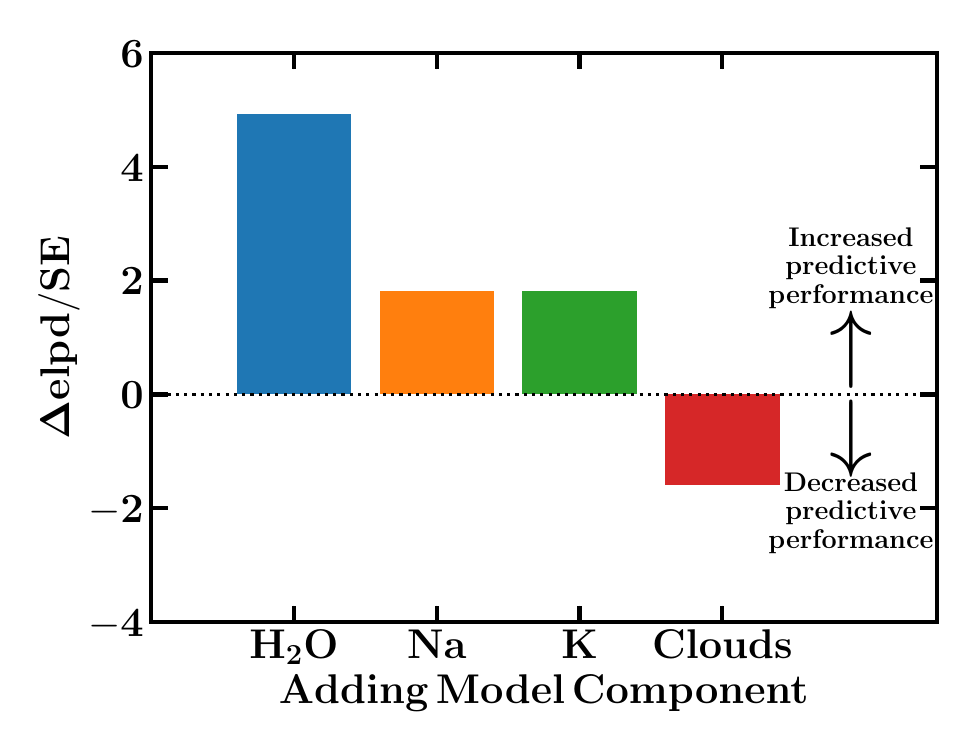}
\caption[]{Predictive model performance quantified by the sum of the $\Delta$elpd scores (Equation \ref{eq:loo_diff}) over the entire dataset divided by their standard error (SE, Equation \ref{eq:se_diff}) see Table \ref{table:synthetic_summary}. The $\Delta$elpd scores are computed by the difference in elpd score between the reference model and models without H$_2$O (blue), Na (orange), K (green), and inhomogeneous clouds and hazes (red). Positive values indicate an increase in the predictive performance of the reference model due to the addition of the specific model component. On the other hand, negative values indicate a decreased predictive performance of the model due to the inclusion of the specific model component.} \label{fig:synthetic_standard_error}
\end{figure} 

In agreement with intuition, the top left panel of Figure \ref{fig:synth_model_loo_comparison} shows that the spectral signatures of H$_2$O in the near-infrared are responsible for the better performance of our model in the HST-WFC3 region ($1.1-1.7\mu m$) of our synthetic observations. Similarly, the top right and bottom left panels of Figure \ref{fig:synth_model_loo_comparison} show that the models with Na and K absorption improve the model performance near $0.59\mu$m and $ 0.77 \mu$m, respectively. Here, and unlike the H$_2$O case for which the largest difference in elpd scores closely follow the influence of the broadband molecular features, the largest difference in elpd score follows the points near the sharp Na and K doublets. Finally, the presence of inhomogeneous clouds and hazes do not improve the model performance significantly and have the smallest difference in elpd scores per data point of all models considered. 

The difference in elpd scores as shown in Figure \ref{fig:synth_model_loo_comparison} allows us to diagnose the performance of the models on a per-data-point level and understand the regions of the data where our model may be under-performing relative to other models. A smaller difference in elpd scores indicates a similar performance between the models considered. Additionally, this metric allows us to infer which subset of the data are responsible for the preference of one model over the other, e.g., infrared HST-WFC3 observations are responsible for the preference of models with H$_2$O absorption.

\begin{deluxetable*}{l|cccccccc}
\tablecaption{Summary Statistics of Retrieval on Synthetic Observations 
\label{table:synthetic_summary}
}
\tablecolumns{9}
\tablehead{
 \colhead{Model}  & \colhead{$\chi^2$} & \colhead{DOF} & \colhead{p-value}  &Log Evidence & \colhead{DS} & \colhead{BIC} & \colhead{elpd Score} & \colhead{$\Delta$elpd Score (SE) }
 }
\startdata
Reference   & $0.90$  & $37$  & $6.48\times10^{-1}$  & $366.4$  & Ref. & $67.65$ & $377.7$ & Ref. \\
No H$_2$O   & $19.47$  & $38$  &  $\rightarrow 0$     & $7.3$ & $27\sigma$  &$770.50$ & $6.0$ & $371.7~(75.5)$ \\
No Na       & $1.76$  & $38$  & $2.53\times10^{-3}$  & $348.2$  &  $6\sigma$  & $97.65$&  $358.8$ & $18.9~(10.4)$ \\
No K        & $0.95$  & $38$  & $5.57\times10^{-1}$  & $365.6$  &  $2\sigma$  & $66.74$ & $376.6$ & $1.2~(0.6)$ \\
Clear atmosphere & $0.81$  & $41$  & $7.98\times10^{-1}$  & $367.4$ &  N/A &  $52.44$ & $378.2$ & $-0.5~(0.3)$ \\
\hline
\enddata
\tablecomments{DS means Detection Significance. The DS is defined for a pair of models with a ratio of evidences (e.g., Bayes factor) $\geq 1$ \citep[e.g.,][]{Trotta2008}. N/A means that the DS is not defined since the reference model has a smaller evidence than the model being tested, e.g., B$<$1. The clear atmosphere model is preferred over the reference model at $2\sigma$. The p-value for the model without H$_2$O is too small and tends to zero.}
\end{deluxetable*}

Finally, we can compare the difference in elpd scores between two or more pairs of models. Figure \ref{fig:synthetic_standard_error} shows the total difference in elpd scores between the reference model and the less complex models in units of their standard error (i.e., Equations \ref{eq:loo_diff} normalized by Equation \ref{eq:se_diff}). Figure \ref{fig:synthetic_standard_error} shows that the model component responsible for the largest increase in the predictive performance of our reference model is H$_2$O absorption. Relative to no improvement in model performance (i.e., a elpd difference of zero), the inclusion of H$_2$O absorption improves the out of sample predictive performance of the model by $\sim 5$ standard errors. This improvement is followed by Na ($\sim2$ standard errors) and K ($\sim2$ standard errors). On the other hand, the inclusion of inhomogeneous clouds and hazes decreased the predictive performance of the model at $\sim1.5$ standard errors.

Table \ref{table:synthetic_summary} summarizes these results alongside other common model comparison metrics for our analysis of these synthetic observations. Despite the similarities between the Log Evidence and the elpd score, or the $\Delta$elpd score divided by its standard error and the `detection significance' in $\sigma$--units, we caution against interpreting them as interchangeable quantities as explained in Section \ref{sec:loo}. Importantly, we compute the standard error (i.e., the uncertainty) of the reported $\Delta$elpd score, unlike the detection significance which is rarely quantified by its numerical uncertainty\footnote{By numerical uncertainty, we mean the uncertainty in the model evidence due to the algorithm used to estimate this quantity, e.g., the log evidence uncertainty as determined by the nested sampling output. Accounting for these numerical uncertainties can change the reported detection significance by up to $\sim0.5\sigma$ in some cases.}(see Section \ref{sec:loo}). 

The information provided by the elpd metric for this atmospheric model for which we know the true input agrees with our expectations. Using this metric enables us to identify deficiencies in the models as a function of wavelength, e.g., the lack of H$_2$O absorption fails to explain the observations in the near-infrared. Additionally, it provides us a way to quantify the improvement in our models by quantifying the out of sample predictive performance of each model and comparing them. This demonstration on a well understood atmospheric scenario showcases the power of this metric to provide us information at the per-data-point level and highlight the strengths and deficiencies of our models. Besides its application to well understood atmospheric scenarios, elpd can help us diagnose the impact of less understood model considerations such as planet inhomogeneities (e.g., clouds, hazes, and multi-dimensional effects), chemical detections of species with unclear spectroscopic features (e.g., metal hydrides in the optical, or H- opacity), and the effects of instrument systematics (e.g., correlated and non-Gaussian noise).

\section{The case of H$^-$ in HAT-P-41b}
\label{sec:hatp41b}

Having built intuition for the elpd metric, we proceed to implement it to the transmission spectrum of the hot Jupiter HAT-P-41b spanning the UV to infrared wavelengths from 0.2-5.0~$\mu$m. HAT-P-41b is the first hot Jupiter for which significant H$^-$ absorption has been invoked as the most robust explanation for its transmission spectrum by \citet{Lewis2020}. With a retrieved H$^-$ abundance several orders of magnitude larger than thermochemical equilibrium expectations \citep[e.g., ][]{Kitzmann2018b, Parmentier2018}, understanding the robustness and reliability of this inference is paramount. By implementing the elpd metric to the same data and atmospheric setup used to infer the presence of H$^-$, we seek to identify which data points drive this possible detection.

We perform a retrieval following the ``minimal'' model in \citet{Lewis2020}, as explained in Section \ref{subsec:model_setup}, as this is the setup for which H$^-$ was inferred at a $\gtrsim2.6\sigma$ level. We perform the retrieval on the combination of HST WFC3 G141, Spitzer IRAC observations and the HST UVIS G280 data from the jitter decorrelation or systematic marginalization treatments in \citet{Wakeford2020}. We compare our retrieved H$^-$ abundances and isothermal temperatures to those from \citet{Lewis2020} for the same data and model configurations. Furthermore, we compare our derived $\chi^2_{\nu,\text{min}}$ and difference in Bayesian evidence, translated to a `detection significance' using the formalism in \citet{Benneke2013} as explained in \citet{Welbanks2021}, to those reported in \citet{Lewis2020}.

When using the UVIS jitter decorrelation data, our retrieval does not find evidence of H$^-$ absorption ($<2\sigma$). Our derived `detection significance' of 1.6$\sigma$ is lower by more than $1\sigma$ to the reported value of 2.9$\sigma$ by \citet{Lewis2020}. Nonetheless, the retrieved $\log(\text{H}^-)=-9.04 ^{+ 0.66 }_{- 1.17 }$ and isothermal temperature of T$=1008 ^{+ 197 }_{- 120}$~K are consistent with the derived properties by \citet{Lewis2020}. Additionally, our $\chi^2_{\nu,\text{min}}=1.55$ is very close to $\chi^2_{\nu,\text{min}}=1.50$ in \citet{Lewis2020}. The agreement in retrieved properties and $\chi^2_{\nu,\text{min}}$ alongside a disagreement in associated `detection significance' may be the result of dissimilar priors and subtle model assumptions between our work and those employed in \citet{Lewis2020}.

On the other hand, when considering the systematic marginalization HST UVIS G280 observations our retrieval finds a 2.7$\sigma$ model preference for H$^-$ absorption, comparable to the 2.6$\sigma$ for the same model and data from \citet{Lewis2020}. This configuration retrieves $\log(\text{H}^-)=-9.32 ^{+ 0.52 }_{- 0.51 }$ for an isothermal temperature of T$=984^{+ 205}_{- 163}$~K consistent with \citet{Lewis2020}. Likewise, the derived $\chi^2_{\nu,\text{min}}=1.83$ is comparable to the value of $\chi^2_{\nu,\text{min}}=1.72$ from \citet{Lewis2020}. This configuration reproduces the weak-to-moderate evidence of H$^-$ in HAT-P-41b. 

\begin{figure*}
\includegraphics[width=1.0\textwidth]{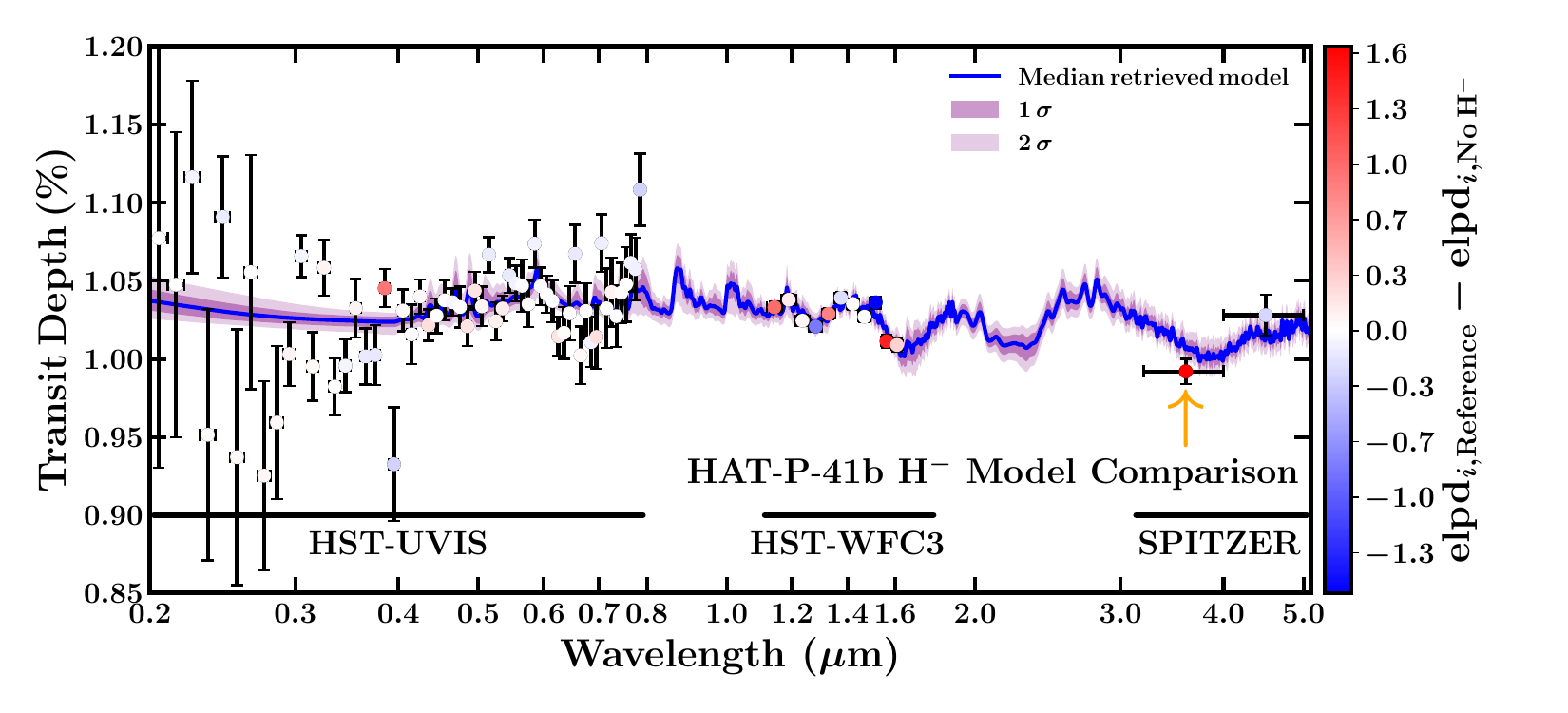}
\caption[]{Retrieved transmission spectrum of HAT-P-41b for the reference ``minimal'' model and HST UVIS G280 observations with systematic marginalization treatment, HST WFC3 G141, and Spitzer IRAC 3.6$\mu$m and 4.5$\mu$m observations from \citet{Wakeford2020} and \citet{Lewis2020}. Data points are color coded by the $\Delta$elpd score between the reference ``minimal'' model and the model without H$^-$ absorption. Redder data points (larger positive $\Delta$elpd score) are better explained by the reference model with H$^-$ absorption while the bluer points (larger negative $\Delta$elpd score) are better explained by the model without H$^-$ absorption. Orange arrow indicates the data point with the highest $\Delta$elpd score (3.6$\mu$m), at a wavelength were bound-free H$^-$ absorption is not expected. The retrievals indicate a 2.7$\sigma$ detection of H$^-$, but removing the data point with the highest $\Delta$elpd score (i.e., Spitzer point at 3.6$\mu$m) removes this preference.} \label{fig:HATP41b_loo}
\end{figure*}

With the reproduction of the weak-to-moderate detection of H$^-$ in HAT-P-41b in mind, we proceed to calculate the approximated PSIS elpd values, as explained in Section \ref{sec:loo}, to quantify out of sample predictive performance for the reference model relative to the model without H$^-$ absorption. Figure \ref{fig:HATP41b_loo} shows the retrieved transmission spectrum and confidence intervals for this retrieval, with the data points color coded by their $\Delta \text{elpd}$ score (Equation \ref{eq:loo_diff}) between the reference model including H$^-$ absorption and the model without H$^-$ absorption. Points with a large positive value (redder points) are better explained by the reference model with H$^-$ (i.e., the reference model has a better out of sample predictive performance), while the large negative values (bluer points) are better explained by the model without H$^-$ absorption (i.e., the reference model has a worse out of sample predictive performance).

The data point with the largest $\Delta \text{elpd}=1.69$ in Figure \ref{fig:HATP41b_loo} is the Spitzer IRAC point at 3.6~$\mu$m, while the point with the lowest $\Delta \text{elpd}=-1.56$ is at $\sim1.52~\mu$m from HST WFC3 G141. Of the four points with the largest $\Delta \text{elpd}$, two are from the HST WFC3 G141 observations (i.e., points at $\sim1.14~\mu$m and $\sim1.56~\mu$m) covering wavelengths at which the spectral contribution of H$^-$ may be expected \citep[see e.g., Figure 7 in][]{Lewis2020}, while two fall outside of this spectral range (i.e., points at $3.6~\mu$m and $0.385~\mu$m). 

The Spitzer IRAC point at 3.6~$\mu$m, highlighted by an arrow in Figure \ref{fig:HATP41b_loo}, is the point with highest $\Delta \text{elpd}$ score. To assess the robustness of the H$^-$ detection in HAT-P-41b, we perform an additional retrieval with the same ``minimal'' model as before but without the data point with the highest $\Delta \text{elpd}$. We find that without the Spitzer IRAC point at 3.6~$\mu$m the reference model with H$^-$ absorption is not preferred ($<2\sigma$) over the model without H$^-$. This means that the 2.7$\sigma$ `detection significance' previously quoted depends on the transit depth at 3.6~$\mu$m being reliable. Unsurprisingly, if both Spitzer photometric points are removed from the retrieval (e.g., only HST UVIS G280 and HST WFC3 G141 observations are considered) we do not find a preference for H$^-$ absorption.

When the Spitzer IRAC point at 3.6~$\mu$m is not considered in the retrieval, the retrieved abundance of H$^-$ is $\log(\text{H}^-)=-9.80 ^{+ 0.57 }_{- 0.83 }$ and the retrieved isothermal temperature is T$=811.69 ^{+ 278.34 }_{- 167.40 }$~K. Additionally, the derived $\chi^2_{\nu,\text{min}}=1.79$ is comparable to that of \citet{Lewis2020}. Overall, while the inferred properties of HAT-P-41b remain largely consistent, removing this single transit depth results in the non-detection of H$^-$ absorption in the atmosphere of HAT-P-41b using current observations. 

Besides the presence of H$^-$, absorption due to H$_2$O has been previously inferred or detected in HAT-P-41b \citep[e.g.,][]{Tsiaras2018, Fisher2018, Lewis2020, Sheppard2021}. Specifically, \citet{Lewis2020} find a $\gtrsim5\sigma$ detection of H$_2$O for their HST WFC3/UVIS, WFC3 G141, and Spitzer IRAC observations, considered in this work, although the model configuration associated with this detection is unclear. For the same set of observations including the systematic marginalization HST UVIS G280, and the ``minimal'' model considered above, our retrieval finds a model preference for H$_2$O absorption at $4.7\sigma$, similar to that reported by \citet{Lewis2020}. 

Figure \ref{fig:HATP41b_loo_h2o} in the Appendix shows the $\Delta \text{elpd}$ for the reference model including H$_2$O absorption and the model without H$_2$O. Following our intuition, the points with the largest positive $\Delta \text{elpd}$ scores, i.e., those better explained by the presence of H$_2$O absorption, fall within the wavelength range of the HST WFC3 G141 observations ($\sim1.1\text{-}1.7\mu$m) where H$_2$O has a prominent absorption feature. The point at $\sim1.38\mu$m has the largest $\Delta \text{elpd}=4.73$ score, almost $3\times$ the highest $\Delta \text{elpd}$ score for the H$^-$ model comparison. 

\begin{figure}
\includegraphics[width=0.47\textwidth]{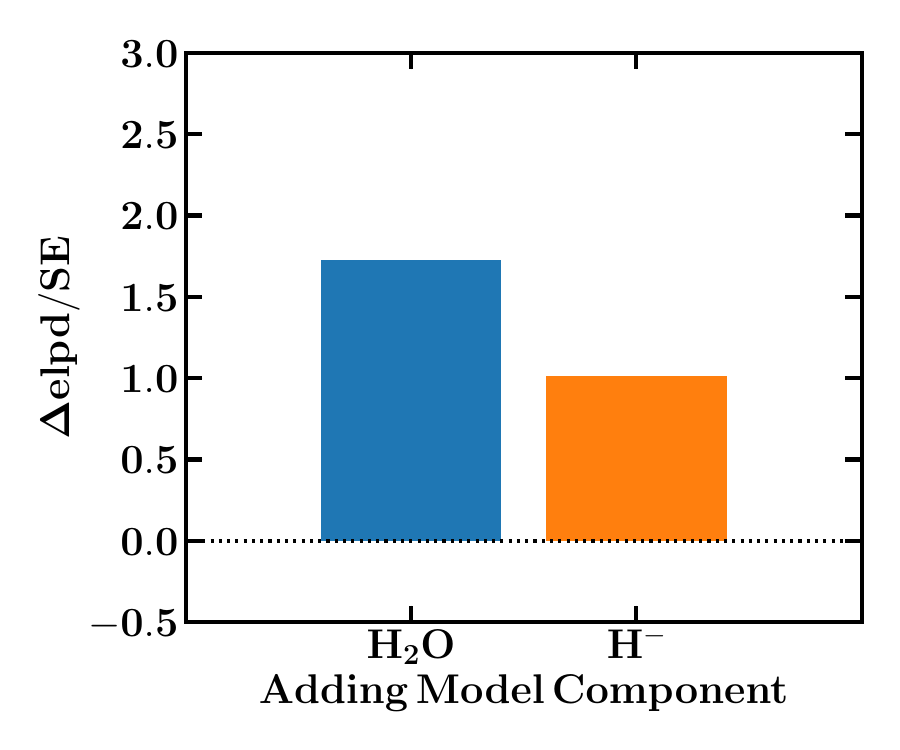}
\caption[]{$\Delta$elpd scores (i.e., predictive model performance) divided by their standard error (SE) for the inclusion of H$_2$O (blue) and H$^-$ (orange) absorption for the reference ``minimal'' model and HST WFC3/UVIS (systematic marginalization), HST WFC3 G141, and Spitzer IRAC observations. Adding H$_2$O absorption into the reference model results in a larger increase in the predictive performance of the model than including H$^-$ absorption.} \label{fig:standard_error_hatp41b}
\end{figure} 

We sum all $\Delta \text{elpd}$ scores over the full data set and calculate the standard error of this difference (Equation \ref{eq:se_diff}) to quantify the overall performance of each of the models considered. Figure \ref{fig:standard_error_hatp41b} shows the overall $\Delta \text{elpd}$ score for each of the models considered divided by their calculated standard error. As before, a larger and positive $\Delta \text{elpd}$ means that adding the model component, e.g., H$_2$O or H$^-$ absorption, increased the predictive performance of the model. We find that including H$_2$O absorption increases the predictive performance of the model by 1.7 standard errors, while including H$^-$ absorption increases the predictive performance by 1.0 standard error. This $\Delta \text{elpd}$ for H$^-$ suggests that H$^-$ absorption does not significantly improve the predictive performance of the model. 

Overall, implementing elpd as a model performance metric on retrievals of transmission spectrum of HAT-P-41b provides complementary per-data-point information, enabling further scrutiny on the possible detection of H$^-$ in the atmosphere of the planet. Particularly, elpd indicates that the weak/moderate ($\gtrsim2.6\sigma$) detection of H$^-$ is contingent on the transit depth at a single data point (i.e., 3.6~$\mu$m). Additionally, the inclusion of H$^-$ on the atmospheric models does not significantly improve the predictive performance of the model. Complementing retrieval studies with elpd analysis, as performed here for the UV to IR spectrum of HAT-P-41b, can provide key information to investigate the robustness of our findings, understand the limits of the data, and contextualize our inferences by the observations. 

\section{Summary and Discussion}
\label{sec:summary_discussion}

Understanding the limitations of our data and our models has become a key need for the robust and reliable interpretation of spectroscopic observations of exoplanet atmospheres. Particularly, as the resolution and precision of our observations improves in the upcoming decade, and atmospheric models include previously neglected physical effects, there is the need for diagnostic metrics that can provide insights into strength and weakness of a model. To that effect we have incorporated elpd to exoplanet atmospheric retrievals.

Unlike conventional model assessment metrics employed in atmospheric retrievals such as the Bayesian evidence or $\chi^2$, elpd provides an assessment of the model performance at the per-data-point level as well as for the model over the entire data set. At its core, elpd provides an estimate for the ability of a model to predict (i.e., explain) each left out data point. By understanding which data points are better explained by each of the models considered, we can quantify the dependency of our inferences on the observations obtained. In other words, elpd provides the tools to contextualize the atmospheric inferences and detections as a function of the observations.  We briefly summarize the main advantages of this tool:

\begin{itemize}
    \item elpd allows each data point to be scored on how well it is predicted by a given model when it is left out of the model fit. This gives us information about the model performance at the per-data-point level.
    \item The elpd score can also be totaled over the entire dataset, giving an estimate of the overall out of sample predictive performance of the model.
    \item The $\Delta$elpd score between different models is an additional metric to assess the need for using more complex models over simpler models with individual components removed. 
    \item elpd helps assess the sensitivity of resulting inferences on specific data points in the dataset.
\end{itemize}

As with any other model comparison metric, elpd has drawbacks and areas that need to be further explored within the context of atmospheric retrievals. These include:

\begin{itemize}
    \item elpd as a metric does not penalize model complexity. Therefore, when assessing models of increasing complexity, elpd should be used alongside metrics like the Bayesian evidence which do penalize model complexity.
    \item elpd score comparison may result in a large number of pairs of models being examined. This could lead, in principle, to a combinatorial explosion of possible pairwise model comparisons. This problem is similar to those of other metrics such as the Bayesian evidence.
    \item elpd should be used alongside existing model assessment and comparison metrics. Together these metrics can provide a holistic picture of the reliability of the data-model interpretation.
\end{itemize}

\subsection{Is there H$^-$ in the Atmosphere of HAT-P-41b?}

As we approach the era of JWST observations, covering wavelengths never probed before with unprecedented precision, it is paramount to determine what constitutes a robust and reliable detection. To date, there is still debate as of what constitutes a clear atmospheric signature, especially for chemical species or atmospheric processes with uncertain spectroscopic signatures. For example, a source of continuum opacity such as H$^-$ does not provide strong spectral features as other prevalent species in exoplanet atmospheres such as H$_2$O, Na, or K. Therefore, previous studies have resorted to calculating the difference in Bayesian evidence for models with and without a model component to quantify its presence, a common practice in the field.

Indeed, as explained in Section \ref{sec:hatp41b}, \citet{Lewis2020} report the tantalizing need for significant H$^-$ absorption in the atmosphere of HAT-P-41b to best explain its transmission spectrum. If confirmed, this would suggest strong chemical disequilibrium in the atmosphere of the planet and would imply the need for new atmospheric models capable of capturing more complex physical effects such as photochemical processes. On the other hand, if this suggestion is invalidated this would highlight the challenges in understanding the limitations of our models and our data. Below we contextualize the additional information obtained by using elpd on the transmission spectrum of HAT-P-41b. 

Our retrieval analysis in Section \ref{sec:hatp41b} can reproduce the weak-to-moderate detection ($\sim2.7\sigma$) of H$^-$ in HAT-P-41b using the same ``minimal'' model and the HST WFC3/UVIS G280 with systematic marginalization, HST WFC3 G141, and Spitzer IRAC observations as \citet{Lewis2020}, by means of Bayesian evidence comparison. However, when using the jitter decorrelation HST WFC3/UVIS G280 and the same ``minimal'' model we obtain a non-significant ($1.6\sigma$) preference for H$^-$ absorption. Our analysis suggests that the presence of H$^-$ is only supported by the combination using the systematic marginalization data.

Our elpd analysis on the retrieval configuration that supports the results from \citet{Lewis2020} provides two main insights. First, the best explained data point by the model including H$^-$ absorption is at a wavelength at which signatures of H$^-$ contribution are not strong. If the data point with the highest $\Delta \text{elpd}$ score is removed from the retrieval analysis, the presence of H$^-$ absorption is no longer preferred by our models. Second, the increase in predictive performance of the models with H$^-$, i.e., how much better does the model with H$^-$ explain the observations relative to the model without H$^-$, is not significant ($\lesssim 1$ SE). In comparison the presence of H$_2$O absorption, the other species inferred/detected by several studies \citep[e.g.,][]{Tsiaras2018,Fisher2018, Lewis2020, Sheppard2021}, is detected at $\sim5\sigma$, better explains the observations at wavelengths where H$_2$O absorption is expected, and increases the predictive performance of the models more significantly ($\sim 2$ SE).

The sensitivity of the H$^-$ detection to the Spitzer IRAC point at 3.6~$\mu$m, may be the result of changes to the continuum level of the spectrum when H$^-$ absorption is included or not in the model. When included, the bound-free H$^-$ absorption provides a continuum in the optical and infared wavelengths (i.e.,$\lesssim2\mu$m) that can help explain the HST UVIS and HST WFC3 G141 observations, but due to its decline in absorption at longer wavelengths allows for a lower transit depth in the 3.6~$\mu$m datapoint. This opacity drop in the mid-infrared is not present when species other than H$^-$, in combination with other model components (e.g., retrieved radius at 10~bar), are used to explain the HST UVIS and HST WFC3 G141 observations, resulting in a slightly higher transit depth at 3.6~$\mu$m. The later is less compatible with the Spitzer IRAC 3.6~$\mu$m observation. This level of model criticism is achievable for the first time thanks to the implementation of Bayesian leave-one-out cross validation.

Therefore, our analysis suggests that H$^-$ is weakly detected in the atmosphere of HAT-P-41b if systematic marginalization treatment of the HST WFC3/UVIS G280 is correct, and the transit depth of the Spitzer IRAC photometric point at 3.6~$\mu$m is precise and accurate. Existing and future observations at these and complementary wavelengths can help verify this inference. For example, the retrieval analysis of \citet{Sheppard2021} using HST STIS observations ($0.3\text{-}1.0~\mu$m) at similar wavelengths does not require significant H$^-$ absorption to explain the transmission spectrum of HAT-P-41b. Future studies could perform a consistent reanalysis of these HST STIS, HST WFC3/UVIS, HST WFC3 G141, and Spitzer IRAC observations to better assess the need for H$^-$ absorption in HAT-P-41b. Finally, future JWST observations with NIRSPEC or NIRCAM could provide important insights in the near-IR where the interaction of electrons and neutral H, not considered by \citet{Lewis2020} or this work, can have significant contributions to the observed spectrum. 

\subsection{Future Considerations}
\label{subsec:future}

For the work in the paper, including the PSIS elpd approximation, we have operated under the assumption of additive white Gaussian noise for all models. Recent work by \citet{Jegug2021} suggests that the presence of more complex and correlated noise of instrument or stellar origin could impact atmospheric parameter estimation. The methods in the paper can be generalized to general correlated Gaussian noise models \citep{Sundararajan2001, Burkner2020}. A future direction of research would be to investigate correlated Gaussian noise models with PSIS-elpd using the Gaussian Process \citep{Rasmussen2005, Ambikasaran2015,celerite1, celerite2} noise model implemented in Aurora \citep{Welbanks2021}.

Further cross-validation schemes specific to exoplanet atmospheric data sets could also be developed in the future  \citep[see e.g.,][for an example of a cross-validation scheme appropriate for time-series data]{Burkner2020}. In these schemes, instead of leaving out one data point all the data points coming from a single instrument could be left out. This particular cross-validation scheme could be used to diagnose systematic instrument effects. Moreover, a cross-validation scheme leaving out several data points clustered around a spectral feature could be explored to differentiate between localized signatures of specific chemical species, and non-localized effects like clouds and hazes.

Bayesian leave-one-out cross validation can help us determine the observational requirements (e.g., resolution, wavelength coverage, signal-to-noise) required for upcoming observational campaigns (e.g., upcoming HST and JWST GO Cycles) and the next generation of large mission concept studies and probe missions \citep[e.g.,][]{Decadal2020}. Particularly, computation of the elpd score using simulated observations can help us understand which observations will be decisive in detecting different atmospheric processes. Furthermore, PSIS (see Section \ref{sec:loo} and Appendix \ref{app:psis_method}) can help us understand the sensitivity of the derived posterior distributions (e.g., the derived atmospheric abundances) to specific observations. 

Finally, although we have demonstrated elpd on low-resolution transmission spectra, the methods introduced here are readily applicable to emission spectra and high-resolution studies. For instance, future studies may investigate the robustness of high-resolution inferences to individual lines or parts of the spectrum. Furthermore, future studies can use elpd to assess the need for additional observations in order to better constrain or confirm previous atmospheric inferences.

\subsection{Concluding Remarks}

We are about to embark into journey of discovery. The imminent high-precision infrared data from JWST \citep{Greene2016}, is bound to revolutionize our understanding of exoplanet atmospheres. These long-awaited observations may help us infer previously unseen chemical species, directly ascertain the 3D properties of exoplanets, and deduce disequilibrium processes that will build towards our search for life in other worlds. These radical discoveries can only be trusted if we acquire the ability to reliably extract the atmospheric properties from an observed spectrum with a thorough understanding of the limits of the data and models. In this context, elpd opens our eyes to a new dimension where we can clearly identify the specific data points upon which our inferences hinge. This new tool helps us revisit our data analysis and check for its robustness, identify the limitations of our models, and plan for the observations required to confirm our discoveries. The diverse and complementary set of tools to assess the robustness and reliability of our discoveries will help us navigate this exploration into the unknown. 

\acknowledgments
{Support for this work was provided by NASA through the NASA Hubble Fellowship grant \#HST-HF2-51496.001-A awarded by the Space Telescope Science Institute, which is operated by the Association of Universities for Research in Astronomy, Inc., for NASA, under contract NAS5-26555. This work was performed using resources provided by the Research Computing at Arizona State University. }

\software{Aurora \citep{Welbanks2021}, Matplotlib \citep{Matplotlib}, MultiNest \citep{Feroz2009, Feroz2013}, Numpy \citep{numpy}, Pandexo \citep{Batalha2017b}, PyMultiNest \citep{Buchner2014}, SciPy \citep{scipy}, Arviz \citep{arviz_2019}.}

\appendix
\restartappendixnumbering
\twocolumngrid 

\section{Pareto Smoothed Importance Sampling Approximation}\label{app:psis}

\subsection{Method}\label{app:psis_method}

Due to the significant computational cost required to calculate $\text{elpd}_{i, \mathcal{M}}$ naively for exoplanet atmospheric models and data sets, we use the Pareto Smoothed Importance Sampling \citep[PSIS;][]{Vehtari2015} approximation outlined in  \citep{Vehtari2017}. The following sections closely follow \cite{Vehtari2015} and \cite{Vehtari2017}. 

Generally, the PSIS approximation works by instead of refitting the model $N$ times, the model is fit to the whole data set and then the posterior samples are re-weighted via importance sampling to estimate the effect of leaving each data point out. The model is only refit when the PSIS approximation is deemed poor. The procedure allows $\text{elpd}_{i, \mathcal{M}}$ to be estimated with significantly less refits of the model and makes computing elpd feasible.   

In general, importance sampling is a Monte Carlo method that allows the expectation of some hard to sample distribution ($f$) to be computed,

\begin{equation}
    \mathbb{E}_{f}[h(\theta)] = \int h(\theta)f(\theta)d\theta.
    \label{eq:IS}
\end{equation}

\noindent Since $f$ is hard to sample from, we wish to use samples from a different distribution ($g$) that is easy to sample from, or has samples readily available, to evaluate Equation \ref{eq:IS}. Providing the ratio $r$ of $f$ and $g$ is known up to some normalization for the samples from $g$ ($\{\hat{\theta}^{s}_{g}\}^{S}_{s=1}$), Equation \ref{eq:IS} can be estimated with importance sampling as,

\begin{equation}
    \mathbb{E}_{f}[h(\theta)] \approx \frac{\sum^{S}_{s=1}r(\hat{\theta}^{s}_{g})h(\hat{\theta}^{s}_{g})}{\sum^{S}_{s=1}r(\hat{\theta}^{s}_{g})}.
    \label{eq:importance_sampling}
\end{equation}

Here, $r(\hat{\theta}^{s}_{g})=f(\hat{\theta}^{s}_{g})/g(\hat{\theta}^{s}_{g})$ is the ratio of $f$ and $g$ up to some normalization constant, and the samples from $g$ are being re-weighted by the importance weights $r(\hat{\theta}^{s}_{g})$.

The standard importance sampling approximation in Equation \ref{eq:IS}, however, is typically unstable due to extreme values of the importance weights dominating the approximation \citep{Vehtari2015}. To remedy this problem PSIS is used instead of standard importance sampling. In PSIS, the raw importance weights ($r(\hat{\theta}^{s}_{g})$) in Equation \ref{eq:IS} are replaced with a set of ``smoothed" weights ($\tilde{r}(\hat{\theta}^{s}_{g})$) which stabilize the approximation. The smoothed weights are calculated by fitting a Pareto distribution \citep{zhang2009} to the tail of the raw weights and replacing the extreme raw weights with draws from the fitted Pareto distribution. 

In addition to stabilizing the importance sampling approximation, PSIS provides a diagnostic on how well the importance sampling approximation will work. Essentially PSIS fails loudly. When the posterior distribution from fitting the full dataset is not a good proposal distribution (i.e., when the posterior distribution changes significantly upon removing the data point in question) the importance sampling will fail. The loud-failing diagnostic is the value of the shape parameter of the fitted Pareto distribution (Pareto $k$). The Pareto $k$ value traces the number of finite moments of the raw weights distribution and therefore the quality of the importance sampling approximation. \cite{Vehtari2015} empirically find that if $k<0.7$, the PSIS approximation is likely to be reliable, whereas $k>0.7$ indicates the PSIS approximation is unreliable and should not be used.

To compute elpd approximately and quickly, PSIS is used in the following way \citep{Vehtari2017}. First the model is fit to the full data set. We then have samples readily available from the posterior distribution where $g =p(\vec{\theta}|\mathcal{D},\mathcal{M})$ in Equation \ref{eq:importance_sampling}. We wish to re-weight these samples using PSIS to approximate Equation \ref{eq:elpd}. To use PSIS, we see by comparing Equations \ref{eq:elpd} and \ref{eq:importance_sampling} that we need to compute the ratio,

\begin{align}
    r(\vec{\theta}) &= \frac{p(\vec{\theta}|\mathcal{D}_{-i},\mathcal{M})}{p(\vec{\theta}|\mathcal{D},\mathcal{M})}\propto  \frac{p(\vec{\theta}|\mathcal{M})p(\mathcal{D}_{-i}|\vec{\theta},\mathcal{M})}{p(\vec{\theta}|\mathcal{M})p(\mathcal{D}| \vec{\theta},\mathcal{M})}\\&= \frac{p(\mathcal{D}_{-i}|\vec{\theta},\mathcal{M})}{p(\mathcal{D} | \vec{\theta},\mathcal{M})}.
    \label{eq:posterior_ratios}
\end{align}

Here, the proportionality arises from Bayes theorem which states that the posterior equals the prior $\times$ likelihood up to a normalization constant (i.e., the evidence). All models in this work have a white Gaussian noise model which factorizes over the data points which means the ratio simplifies to,

\begin{equation}
    r(\vec{\theta}) \propto \frac{1}{p(\mathcal{D}_{i} | \vec{\theta},\mathcal{M})}.
    \label{eq:raw_weights}
\end{equation}

Putting everything together and denoting $\hat{\theta}^{s}$ as one of $S$ samples from the full posterior $p(\vec{\theta}|\mathcal{D}, \mathcal{M})$ we have,

\begin{equation}
    p(\mathcal{D}_{i}|\mathcal{D}_{-i}, \mathcal{M})\approx \frac{\sum^{S}_{s=1}\tilde{r}(\hat{\theta}^{s})p(\mathcal{D}_{i}|\hat{\theta}^{s}, \mathcal{M})}{\sum^{S}_{s=1}\tilde{r}(\hat{\theta}^{s})}.
    \label{eq:PSIS}
\end{equation}

\noindent For each of the left of data points the raw weights are calculated via Equation \ref{eq:raw_weights} and are then smoothed. If $k<0.7$ we deem PSIS reliable and use it to calculate $\text{elpd}_{i, \mathcal{M}}$ via Equation \ref{eq:PSIS}. For each data point where smoothing results in $k>0.7$ we do not use the PSIS approximation. Instead we perform a full Baysian refit with the data point left out and evaluate $\text{elpd}_{i, \mathcal{M}}$ exactly via Equation \ref{eq:elpd_exact}. Overall, this procedure allows the computation of all $\text{elpd}_{i, \mathcal{M}}$ terms with only a small number of expensive refits of the model.

\subsection{Validating the PSIS Approximation}\label{app:psis_validation}

\begin{figure*}
\includegraphics[width=1.0\textwidth]{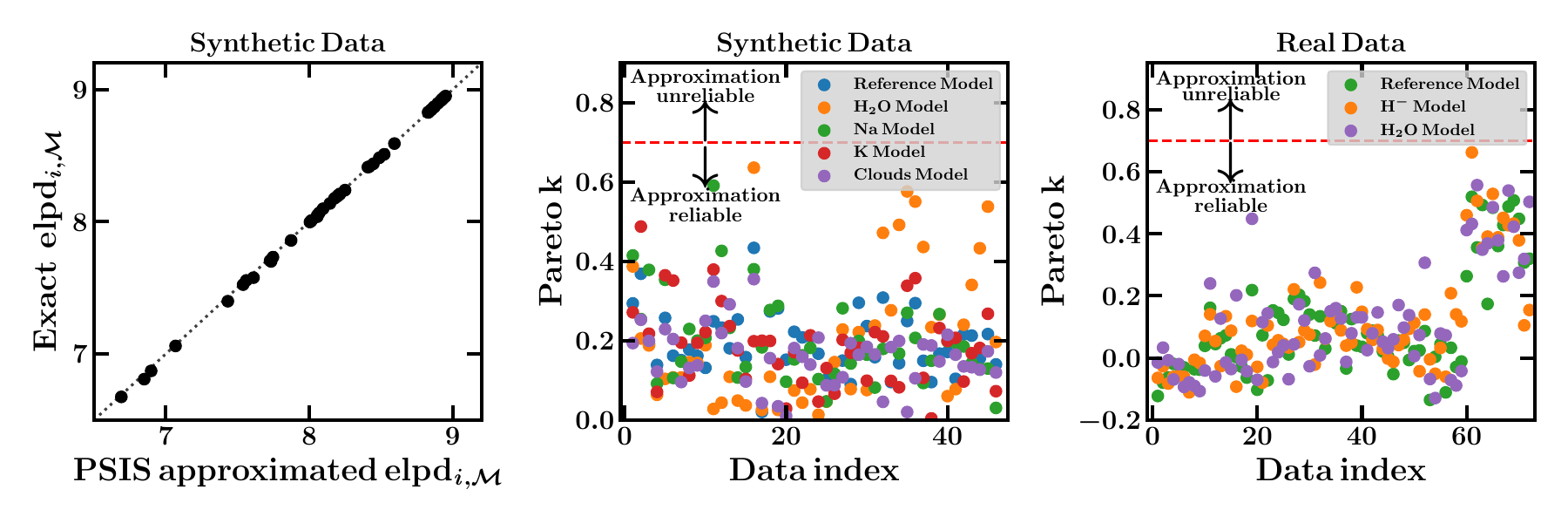}
\caption[]{Left: Exact elpd terms (Equation \ref{eq:elpd_exact}) against PSIS approximated elpd terms (Equation \ref{eq:PSIS}) for the synthetic data scenario described in Section \ref{sec:simulated_data}. The exact terms we calculated by refitting the model $N$ (46) times with each data point left out in turn. One-to-one correspondence is shown as a dashed line. The PSIS approximated elpd terms are in good agreement with the exact terms. Middle: Pareto $k$ values for the PSIS approximation for each data point and for each model considered in Section \ref{sec:simulated_data}. Reference model is shown in blue, while models without H$_2$O, Na, K, or clouds are shown in orange, green, red, and purple, respectively. Data point number is ordered by increasing wavelength. Right: As middle panel but for the real data in Section \ref{sec:hatp41b} and the ``minimal'' model used as reference (green points), the model without H$^-$ absorption (orange points), and the model without H$_2$O absorption (purple points). For the synthetic and real observations, all Pareto $k$ values are $<0.7$ indicating that the PSIS approximation is reliable for all left out data points. This means all elpd terms were computed with PSIS and only one model fit had to be performed.} \label{fig:PSIS_LOO}
\end{figure*}

In this section we assess the accuracy of the PSIS approximation for an exoplanet atmospheric model. Specifically, we use the atmospheric model explained in Section \ref{subsec:model_setup} considering 3 chemical absorbers, an isothermal atmosphere, an the possibility of inhomogeneous clouds and hazes. First we calculate the elpd score exactly by performing independent retrievals in which each individual spectroscopic data point is removed in turn and remains `unseen' (i.e., $N=46$ refits of the model or retrievals). The resulting exact $\text{elpd}_{i}$ terms are evaluated for each data point using Equation \ref{eq:elpd_exact}. Then, we calculate the same $\text{elpd}_{i}$ terms using the PSIS approximation (Equation \ref{eq:PSIS}) The resulting approximated PSIS and exact $\text{elpd}_{i}$ are shown Figure \ref{fig:PSIS_LOO}, along with the Pareto $k$ values from the PSIS approximation.

Figure \ref{fig:PSIS_LOO} shows the approximated PSIS elpd scores and exact elpd scores lie along a unity line and are in agreement with one another. The middle panel of Figure \ref{fig:PSIS_LOO} shows the Pareto $k$ values for the approximation, with all of them below the empirical value of $<0.7$. As a result, the PSIS approximation only required one refit of the model compared to the $46$ refits required to compute all the elpd terms exactly, and still provided accurate results. Overall, the PSIS approximation was accurate and saved $\approx4,000$ CPU hours of computation in this case, i.e., the PSIS approximation took $\approx1/N=1/46$ of the computation of the exact calculation. For completeness, we include in the right panel of Figure \ref{fig:PSIS_LOO} the Pareto $k$ values for the approximations employed in Section \ref{sec:hatp41b}.

\section{Auxiliary retrieval information}\label{app:priors}

\begin{figure*}
\includegraphics[width=1.0\textwidth]{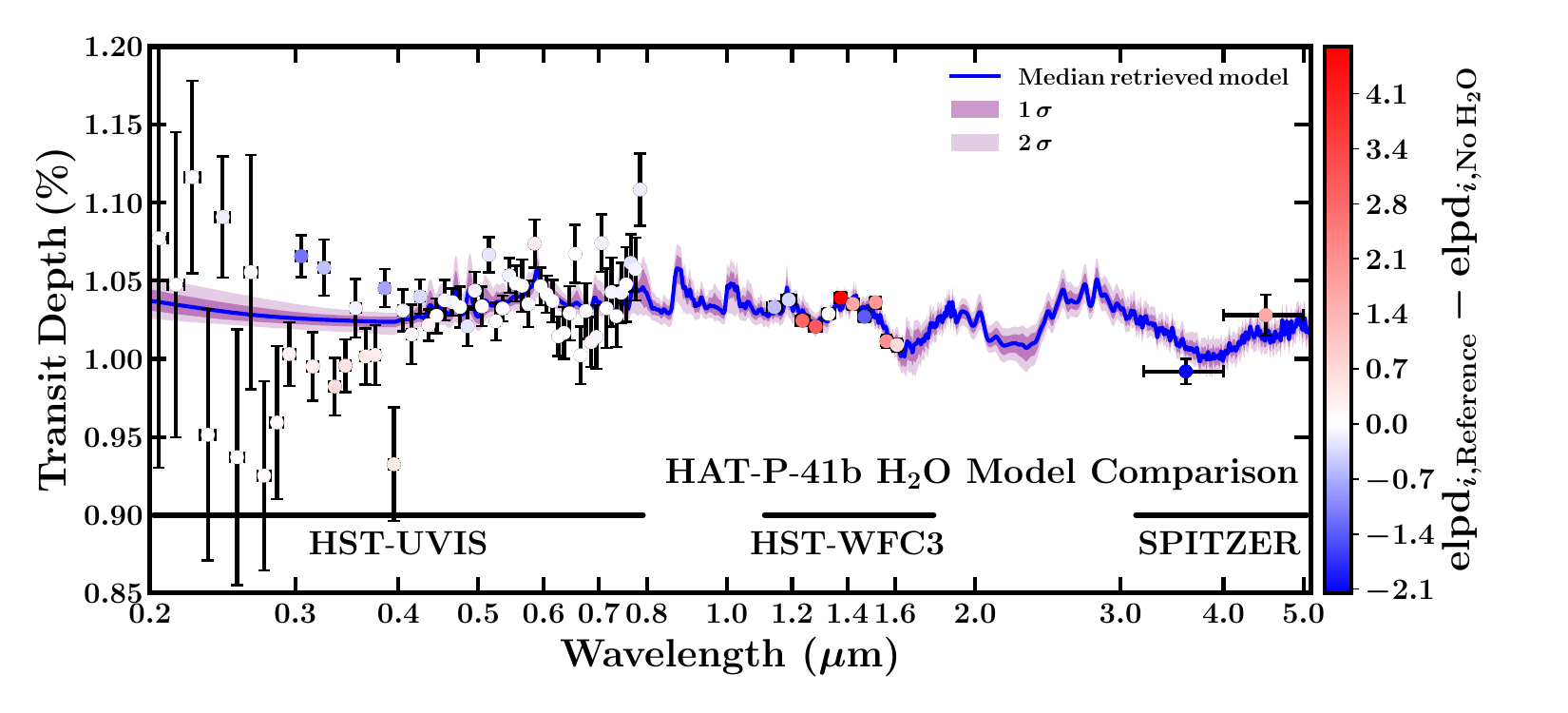}
\caption[]{As Figure \ref{fig:HATP41b_loo} but for the $\Delta$elpd score between the reference ``minimal'' model and the model without H$_2$O absorption. Redder data points are preferentially explained by H$_2$O absorption. The data points with a larger $\Delta$elpd score (redder data points) indicate a preference for H$_2$O absorption, and fall at wavelengths near the $\sim1.4\mu$m H$_2$O feature as expected.} \label{fig:HATP41b_loo_h2o}
\end{figure*}

Table \ref{table:priors} shows the priors used for the retrievals in Sections
\ref{sec:simulated_data} and \ref{sec:hatp41b}. Figure \ref{fig:HATP41b_loo_h2o} shows the retrieved transmission spectrum of HAT-P-41b on the HST UVIS/G280 systematic marginalization, HST WFC3 G141, and Spitzer IRAC observations from \citet{Wakeford2020} and \citet{Lewis2020}. The data points are color coded by their $\Delta \text{elpd}$ score between the reference ``minimal'' model with H$_2$O absorption and the model without H$_2$O absorption. A larger positive value (redder point) indicates that the point is better explained by the presence of H$_2$O absorption (i.e., better out of sample predictive performance by the reference model) while a larger negative value (bluer point) indicates that the datum is better explained by the model without H$_2$O absorption  (i.e., worse out of sample predictive performance by the reference model). As expected, the points with the largest $\Delta \text{elpd}$ score lie within the HST WFC3 observations (e.g., $\sim1.1\text{-}1.7\mu$m) where H$_2$O has a prominent absorption feature.

\begin{deluxetable*}{c|c|c|c|c}
\tablecaption{Parameters and Priors for Retrievals in this Work
\label{table:priors}}
\tablecolumns{4}
\tablewidth{\columnwidth}
\tablehead{
\colhead{Parameter} & \colhead{Prior Distribution} & \colhead{Synthetic Observations} & \colhead{HAT-P-41b Minimal Model}
}
\startdata
$X_i$ & Log-uniform & $10^{-12}$--$10^{-0.3}$ &$10^{-12}$--$10^{-1.0}$\\ \hline
$T_0$ & Uniform &  $800$--$2000$ K   & $400$--$2200$ K \\ \hline
$R_{\rm p, \, 10 \, bar}$ & Uniform & N/A &  $1.4$--$1.8$~$R_{\rm J}$ \\ \hline
$P_{\rm ref. }$ & Log-uniform &  $10^{-6}$--$10^{2}$ bar & N/A\\ \hline
$P_{\rm cloud }$ & Log-uniform &  $10^{-6}$--$10^{2}$ bar & N/A\\ \hline
$a$ & Log-uniform & $10^{-4}$--$10^{10}$ & N/A \\ \hline
$\gamma$ & Uniform & $-20$--$2$  & N/A \\ \hline
$\phi_{\mathrm{clouds+hazes}}$ & Uniform & $0$--$1$ & N/A
\enddata 
\tablecomments{N/A means that the parameter was not considered in the model by construction.}
\end{deluxetable*}

\bibliographystyle{aasjournal}
\bibliography{biblio}

\end{document}